\documentclass{PoS}

\title{Bounds on free energy in QCD}

\ShortTitle{Bounds on free energy in QCD}

\author{{Axel Maas}\thanks{Supported by the DFG under grant number MA 3935/5-1.}\\
        Institute for Theoretical Physics,
              Friedrich-Schiller-University Jena, Max-Wien-Platz 1, D-07743 Jena, Germany\\
        E-mail: \email{axelmaas@web.de}}

\author{\speaker{Daniel Zwanziger}\\
        Physics Department, New York University, 4 Washington Place, New York, NY 10003, USA\\
        E-mail: \email{dz2@nyu.edu}}

\abstract{We derive some exact bounds on the free energy $W(J)$ in QCD, where $J_\mu^b$ is a source for the gluon field $A_\mu^b$ in the minimal Landau gauge, and $W(J)$ is the generating functional of connected gluon correlators.  Among other results, we show that for a static source $J(x) = h$ the free energy vanishes, $W(h) = 0$, together with its first derivative,  ${ \partial W(h) \over \partial h} = 0,$  for all $h$, no matter how strong.  Thus the system does not respond to a static color probe.  We also present numerical evaluations of the free energy $W(J)$ and find that the bounds are well satisfied and in fact undersaturated.}

\FullConference{Xth Quark Confinement and the Hadron Spectrum,\\
		October 8-12, 2012\\
		TUM Campus Garching, Munich, Germany}

\begin{document}

 \def\beq{ \begin{equation}}
 \def\eeq{\end{equation}}
 \def\beqa{ \begin{eqnarray}}
 \def\eeqa{  \end{eqnarray}}
\newcommand{\p}{\partial}

\section{Introduction}

We shall be concerned with the Euclidean correlators of gluons in QCD with an $SU(N)$ local gauge symmetry that are fixed to the minimal Landau gauge.  These are the fundamental quantities in quantum field theory.

The minimal Landau gauge is obtained by minimizing the Hilbert square norm
\beq
|| A ||^2 = \int | A_\mu^b(x) |^2 d^4x,
\eeq  
to some local minimum (in general not an absolute minimum) with respect to local gauge transformations $g(x)$.  These act according to ${^g}A_\mu = g^{-1} A_\mu g + g^{-1} \p_\mu g$.  At a local minimum, the functional $F_A(g) \equiv ||{^g}A||^2$ is stationary and its second variation is positive.  It is well known that these two properties imply respectively that the Landau gauge (transversality) condition is satisfied, $\p \cdot A = 0$, and that the Faddeev-Popov operator is positive {\it i. e.} $(\omega, M(A) \omega) \geq 0$ for all $\omega$.  Here the Faddeev-Popov operator acts according to $M^{ac}(A) \omega^c = - \p_\mu D_\mu^{ac}(A) \omega^c$, the gauge covariant derivative is defined by $D_\mu^{ac}(A) \omega^c = \p_\mu \omega^a + f^{abc} A_\mu^b \omega^c$, and the coupling constant has been absorbed into $A$.  Configurations $A$ that satisfy these two conditions are said to be in the (first) Gribov region \cite{Gribov:1977wm} which we designate by $\Omega$.  It is known that in general there are more than one local minimum of $F_A(g)$, and we do not specify which local minimum is achieved.  This gauge is realized numerically by minimizing a lattice analog of $F_A(g)$ by some algorithm, and the local minimum achieved is in general algorithm dependent.

The analytic bounds which we shall obtain follow from the restriction of the gauge-fixed configurations to the Gribov region $\Omega$, and are the same whether the gluons are coupled to quarks as in full QCD, or not, as in pure gluodynamics.  In fact the same bounds hold for other gauge bosons with $SU(N)$ gauge symmetry, for example, in the Higgs sector, provided only that the gauge-fixing is done to the minimal Landau gauge.  The numerical results we shall present will be for pure gluodynamics in $SU(2)$ gauge theory. 

\section{General bounds on free energy}

In the minimal Landau gauge, the free energy $W(J)$ is defined by
\beq
 \exp W(J) \equiv \int_\Omega dA \ \rho(A) \ \exp(J, A).\label{freeenergy}
 \eeq
Here the quark degrees of freedom (if present) are integrated out.  The Euclidean probability $\rho(A)$ includes the Yang-Mills action, the gauge-fixing factor $\delta(\p \cdot A)$, the Faddeev-Popov determinant, and possibly the quark determinant.  We shall use only the properties $\rho(A) \geq 0$ and $\int dA \ \rho(A) = 1$.  The source term $J_\mu^a(x)$ is real and is taken to be transverse $\p \cdot J = 0$ without loss of generality because $A$ is identically transverse.  The free energy per unit Euclidean volume, $w(J) = W(J)/V,$ is the generating functional of connected correlators,
\beq
\langle A(x) A(y) ... \rangle_{\rm conn} = {\p \over \p J_x}  {\p \over \p J_y} ... w(J).
\eeq

The general bound is immediate.  From the inequality $(J, A) \leq \max_{A \in \Omega} (J, A) = (J, A_{\rm max})$, where $A_{\rm max}$ is that configuration in $\Omega$ that maximizes $(J, A)$ for fixed $J$, we obtain
\beqa
\exp W(J) & \leq &  \int_\Omega dA \ \rho(A) \  \exp(J, A_{\rm max}) \nonumber \\
& = & \exp(J, A_{\rm max}),
\eeqa
which gives the bound
\beq
W(J) \leq \max_{A \in \Omega} (J, A).
\eeq
Because $\Omega$ is bounded in every direction \cite{Zwanziger:1982}, this bound is finite.  It is not hard to show that the maximum occurs when $A$ lies on the boundary $\p \Omega$ of $\Omega$, and we have the more precise bound on the free energy,
\beq
\label{genbound}
W(J) \leq \max_{A \in \p \Omega}(J, A).
\eeq

The right hand side is linear in $J$,
\beq
\max_{A \in \p \Omega}(h J, A) = h \max_{A \in \p \Omega}(J, A)
\eeq
for $h > 0$.  This linear bound is a characteristic feature for integration over any bounded region, such as the Gribov region.  By contrast, for a free field the free energy is quadratic in $J$,
\beq
W_{\rm free}(J) = (1/2) (J, K^{-1}J),
\eeq
where $K = - \p^2 +  m^2$.  This strongly violates the linear bound (\ref{genbound}) at large $J$.

\section{Bound for a plane-wave source}\label{spw}

We now specialize to a plane wave source,
 \beq
 J_\mu^a(x) = h \cos(k \cdot x_2) \delta^{a3} \delta_{\mu 1},\label{source}
 \eeq
Here $h$ is the analog in a spin theory of an external magnetic field, modulated by a plane wave.  The wave number takes on the values $k = 2\pi n/L$, where $n$ is an integer, and $L$ is the edge of a periodic Euclidean box.  The indices $1$ and $2$ are chosen so $J$ is transverse, $\p_\mu J_\mu = 0$.  For this source, we parametrize the free energy per unit Euclidean volume $V = L^d$ by
 \beq
 w(k, h) \equiv W(h \cos(k \cdot x))/V.
 \eeq
 The gluon propagator is its second derivative at $h = 0$,
\beq
D(k) = (1/2) {\p^2 w(k, h) \over \p h^2}\Big|_{h = 0}.
\eeq

  For a static source, $k = 0$, we have $A_{\rm max}(x) = {\rm const}$, and the Faddeev-Popov operator may be diagonalized by Fourier transform.  In this case we obtain on a finite volume $V = L^d$, the bound
\beq
w(0, h) \leq |h| (2\pi/L).
\eeq  
Thus on an infinite volume, $L \to \infty$, the free energy vanishes, 
\beq
\label{staticb}
 \lim_{L \to \infty} w(0, h) = 0,
\eeq
for a static source of strength $h$ no matter how strong.  Lest it be thought that this is a peculiarity of the mode $k = 0$, we also exhibit a bound at infinite volume, with $k$ finite.  By an explicit calculation one obtains \cite{Zwanziger:1991}
 \beq
 \label{kbound}
 w(k, h) \leq |h k|.
 \eeq
The free energy vanishes in the static limit 
 \beq
 \lim_{k \to 0} w(k, h) = 0
 \eeq
 for all $h$ no matter how strong, in agreement with (\ref{staticb}).  
 
      We also obtain a bound on the ``magnetization" $m(k, h) \equiv {\p w(k, h) \over  \p h}$.  Indeed because $\rho(A)$ is normalized, we have $w(k, 0) = 0$, and so
\beq
w(k, h) = \int_0^h dh' \ m(k, h').
\eeq     
Inserting this into the bound (\ref{kbound}), we obtain, for $k > 0$ and $h > 0$,
\beq
{1 \over h} \int_0^h dh' \ m(k, h') \leq k,
\eeq
which gives
\beq
\lim_{k \to 0} {1 \over h} \int_0^h dh' \ m(k, h') = 0.
\eeq
Note that $m(k, 0) = 0$, and that
${\p m(k, \ h) \over \p h} = {\p^2 w(k, \ h) \over \p h^2} \geq 0$ is positive because $w(k, h)$ is a convex function of $h$.  Consequently $m(k, h)$ is positive, $m(k, h) \geq 0$, and we conclude from the positivity of the integrand and the vanishing of the last integral that the magnetization produced by a static source vanishes,
\beq
\lim_{k \to 0} m(k, h) = 0,
\eeq
for almost all $h$.  Thus, in the minimal Landau gauge, the static static color degree of freedom cannot be excited by applying an external color-magnetic field $h$, no matter how strong.
 
 Finally we note from the vanishing of $ w(0, h) \equiv \lim_{k \to 0} w(k, \ h) = 0$ that, IF $w(0, h)$ were analytic in $h$, then all derivatives of $w(0, h)$ with respect to $h$ would vanish.  In particular the second derivative would vanish, ${\p^2 w(0, \ h) \over \p h^2}|_{h = 0} = 2 D(0) = 0$.  This is the statement that the gluon propagator $D(k)$ vanishes at $k = 0$.  Conversely, if $D(0) > 0$, then $w(0, h)$ is non-analytic in $h$.  Numerical data indicate that $D(0) > 0$ in Euclidean dimension $d = 3, 4$,  \cite{Cucchieri:2007rg, Cucchieri:2007md, Cucchieri:2010xr, Bogolubsky:2007ud, Bogolubsky:2009dc, Bornyakov:2009ug}, while $D(0) = 0$ for $d = 2$ \cite{Maas:2007uv, Cucchieri:2007rg, Cucchieri:2011um}.\footnote{The vanishing of $D(0)$ in Euclidean dimension $d = 2$ is proven in \cite{Zwanziger:2012},}  Thus non-analyticity of $w(0, h)$ is implied by the lattice data in Euclidean dimension $d = 3, 4$. 
 
\section{Numerical results for the free energy}

A numerical measurement using a Monte Carlo approach of a current-dependent free energy (\ref{freeenergy}) with a current in a fixed gauge is non-trivial, because (yet) no efficient algorithm for Monte Carlo updates in a fixed gauge is known, though first proposals exist \cite{vonSmekal:2008es}. To circumvent this problem, here reweighting will be used. In this case, the lattice configurations are obtained at zero external current. The current-dependent free-energy $W$ is then measured by
\begin{equation}
\exp(W\left[J\right])=\left\langle\exp\left(\int JA\right)\right\rangle,\label{wsubstitute}
\end{equation}
\noindent which will be performed in the minimal Landau gauge, see \cite{Maas:2011se} for a review. Since the measured quantity is an exponential in the fields, the standard importance sampling cannot be expected to be accurate for large $J$. It will fail at the latest when the source term becomes comparable with the action itself. Of course, it cannot be excluded that already a small value of $J$ upsets the importance sampling significantly.

The situation can be a bit better estimated when using the fixed source term (\ref{source}). Using that the gauge field in lattice units is bounded by one, the source term is of maximum size
\begin{equation}
\max\int dx JA=Vh\label{jbound}
\end{equation}
\noindent The maximum expectation value of the Wilson action used here is given by
\begin{equation}
\max S=\frac{\beta V d(d-1)}{2}\langle P\rangle\nonumber,
\end{equation}
\noindent where $d$ is the number of space-time dimensions, $\beta=4/g^2$, and $\langle P\rangle$ is the plaquette expectation value, i.\ e.\ the free energy per unit volume. Thus, the maximum $h$ possible is
\begin{equation}
h\lesssim \frac{\beta d(d-1)\langle P\rangle}{2}\label{hmax},
\end{equation}
\noindent which still permits rather large $h$, under the assumption that going close to this limit will not distort the results too severely. In the following some preliminary results from such an evaluation will be presented. More results will be presented elsewhere \cite{unpublished}. These will be given for two, three, and four dimensions, as it has been found that gauge-dependent quantities depend significantly on the dimensionality \cite{Maas:2011se}.

\begin{figure}
\includegraphics[width=\linewidth]{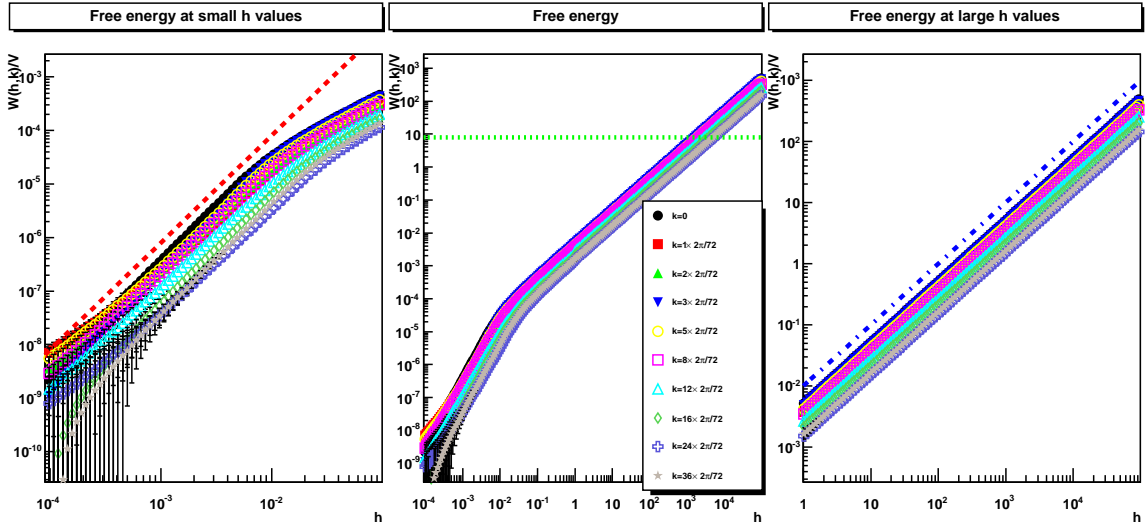}\\
\caption{\label{wres}The free energy density $W(h,k)/V$ in lattice units as a function of $h$ for various values of $k$ is shown in the middle panel. The right and left panels show magnifications of the regions at large and small $h$, respectively. The value of $h$ is limited by (\protect\ref{hmax}), indicated by the green dashed line in the top panel. In the right panel the blue dashed line is linear in $h$, while the red dashed line in the left panel is proportional to $h^2$. The three-dimensional lattice has volume of 72$^3$ lattice sites at $\beta=3.73$, i.\ e.\ (14.4 fm)$^3$ and $a=0.2$ fm.}
\end{figure}

With the source (\ref{source}), the functional $W$ becomes a function of the two independent variables $k$ and $h$. Since $k$ is a lattice momentum, it can have only discrete values, while $h$ is a continuous variable. For $k$, at most ten different values have been used, depending on the lattice size. An example for $W$ is shown in figure \ref{wres}. It is immediately visible that $W$ depends polynomial on $h$ at large values of $h$. Since this sets in already some orders of magnitude below the reliability limit (\ref{hmax}), this is likely a genuine effect.

\begin{figure}
\includegraphics[width=0.333\linewidth]{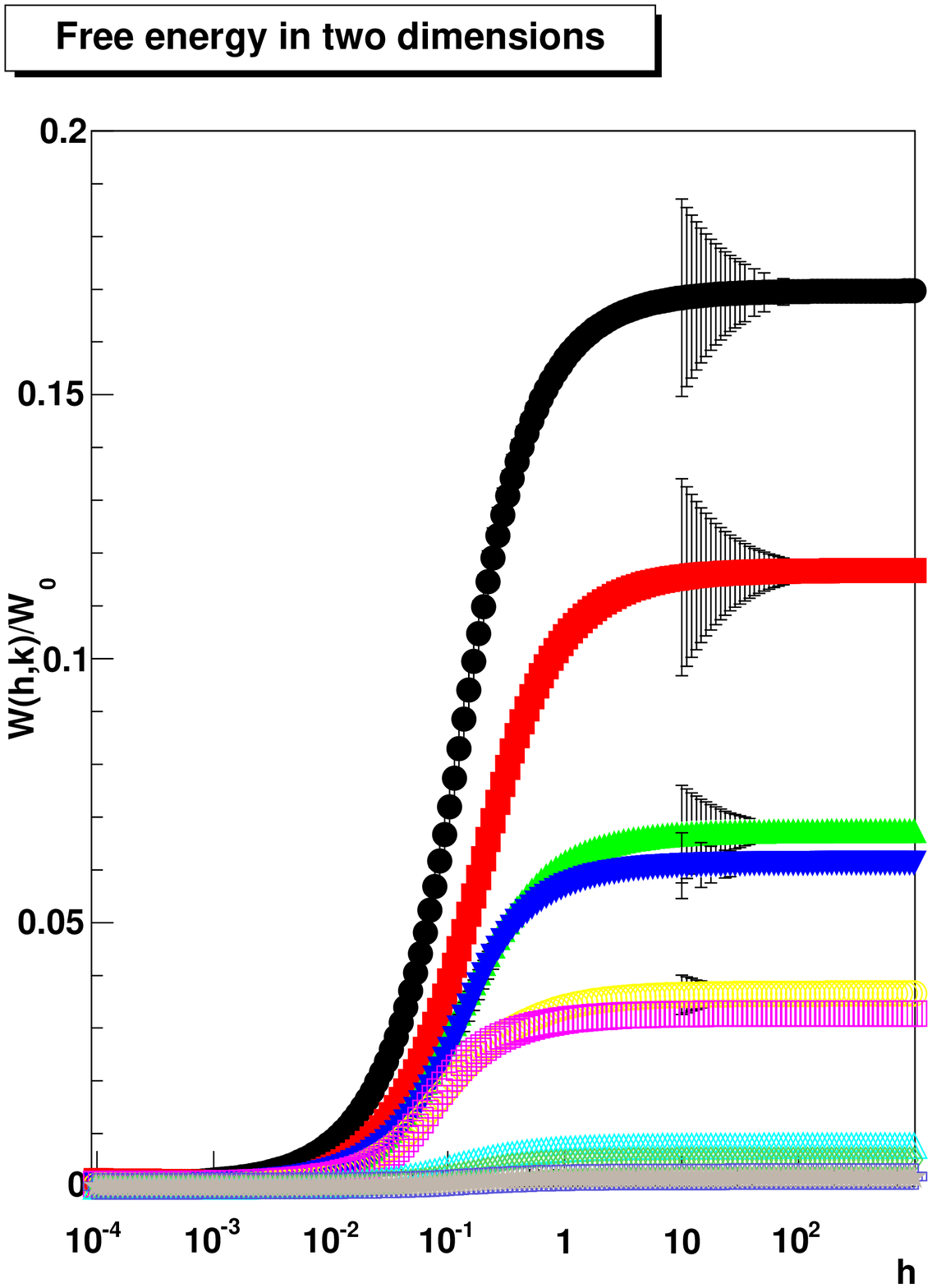}\includegraphics[width=0.333\linewidth]{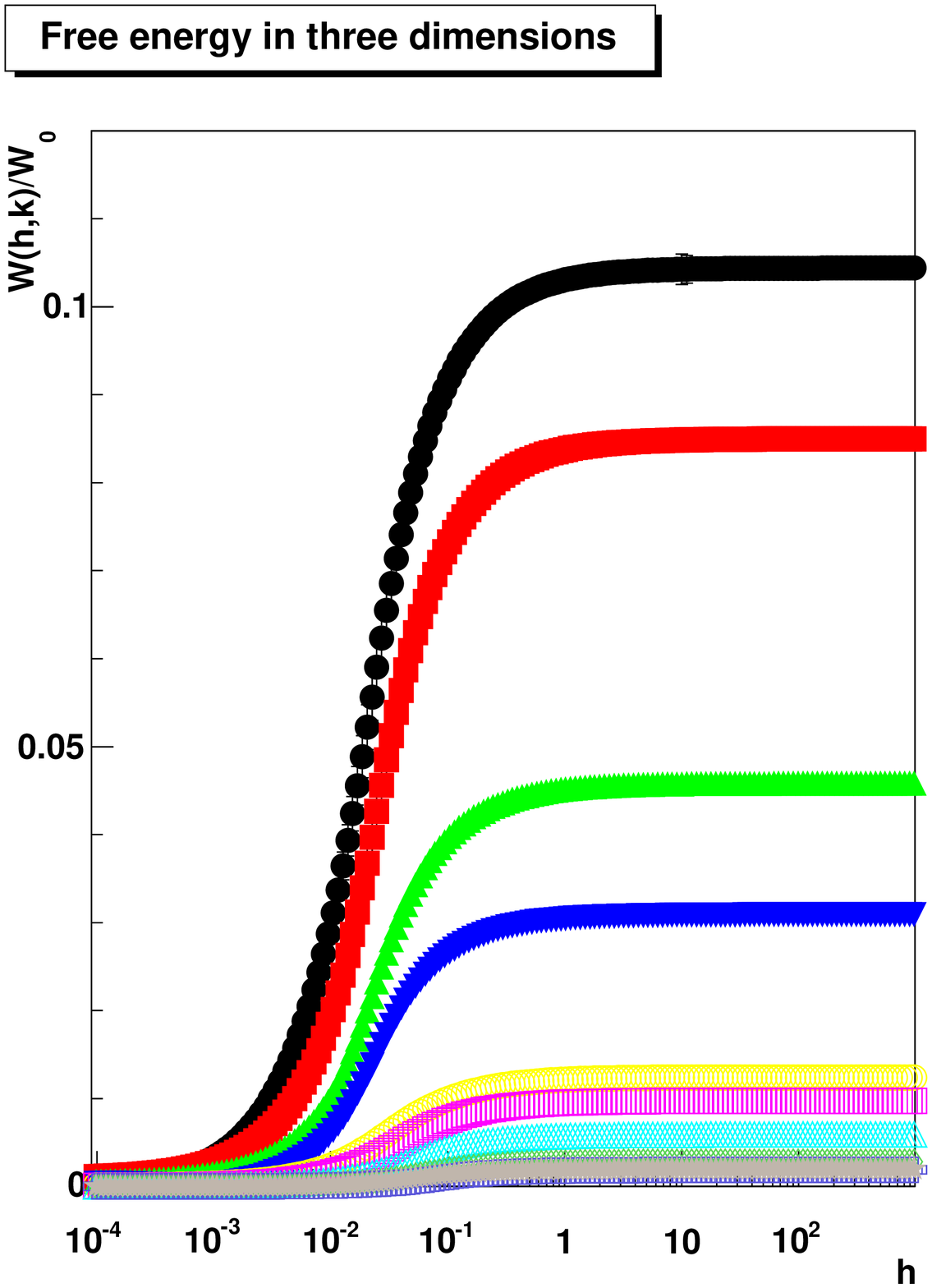}\includegraphics[width=0.333\linewidth]{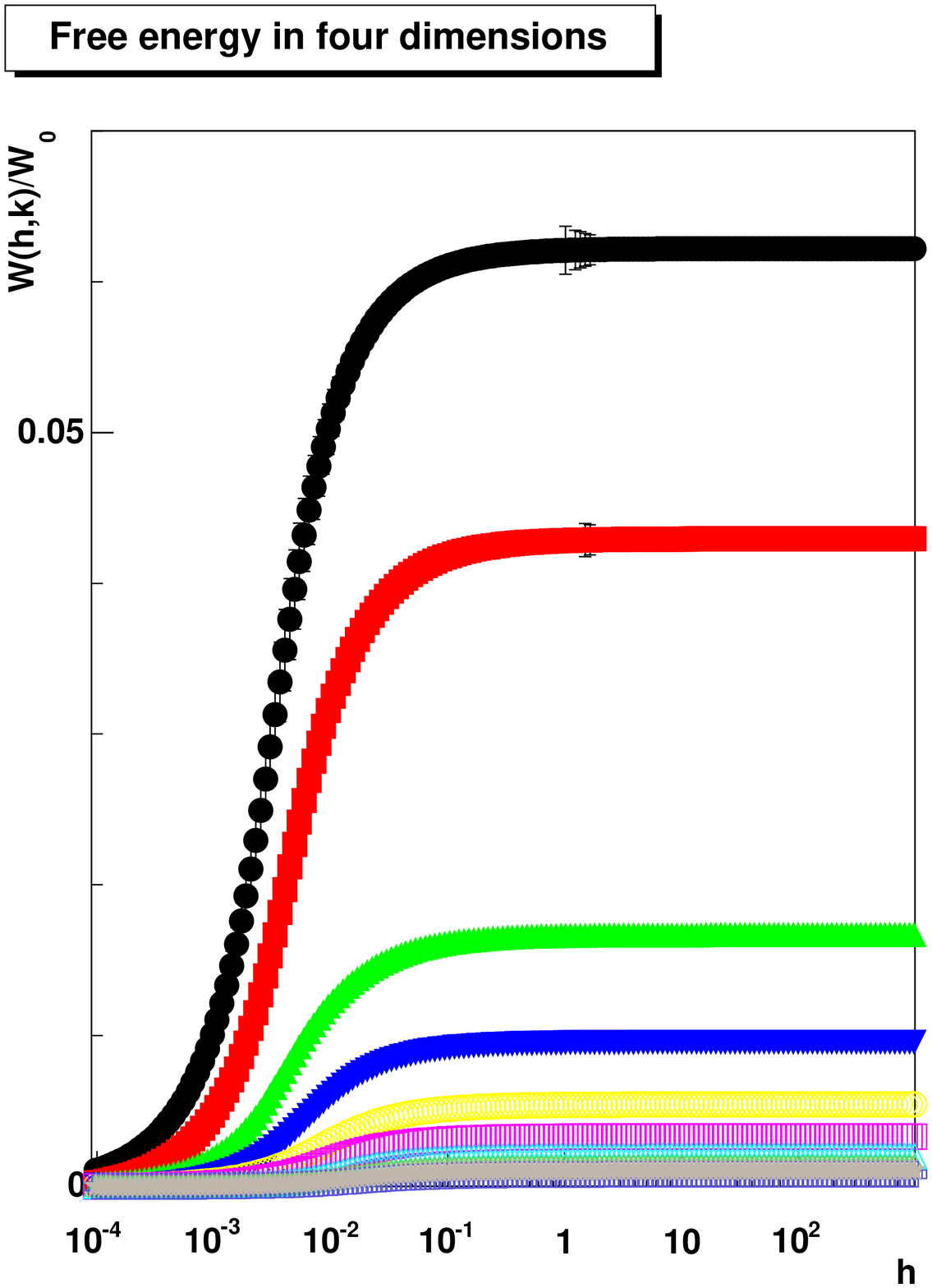}
\caption{\label{wnorm}The normalized free energy $W/W_0$ as a function of $h$ for two (left panel, 240$^2$/(48 fm)$^2$ at $\beta=7.99$/$a=0.2$ fm), three (middle panel, 36$^3$/(7.2 fm)$^3$ at $\beta=3.73$/$a=0.2$ fm), and four (right panel, 24$^4$/(4.8 fm)$^4$ at $\beta=2.221$/$a=0.2$ fm) dimensions. See \cite{unpublished} for an explanation of the somewhat unusual error distribution. The different colors and symbols refer to increasing values of $k$, with values depending on the lattice parameters, but generically the large the value of $W/W_0$, the smaller $k$ and the largest values always correspond to $k=0$.}
\end{figure}

To identify the slope, figure \ref{wnorm} shows $W/W_0$, where $W_0$ is $\sqrt{2}\pi N^{d-1} h$ for $k=0$ and $N^d/\sqrt{2} h k$ otherwise. This quantity shows the linear dependency on both $h$ and $k$ (\ref{kbound}) expected at large $h$, and any correction is found to be smaller than ${\cal O}(\ln(h))$ already substantially below the limit (\ref{hmax}). However, the bound $W_0$ is not saturated, and a prefactor smaller than one remains.

\begin{figure*}
\includegraphics[width=\linewidth]{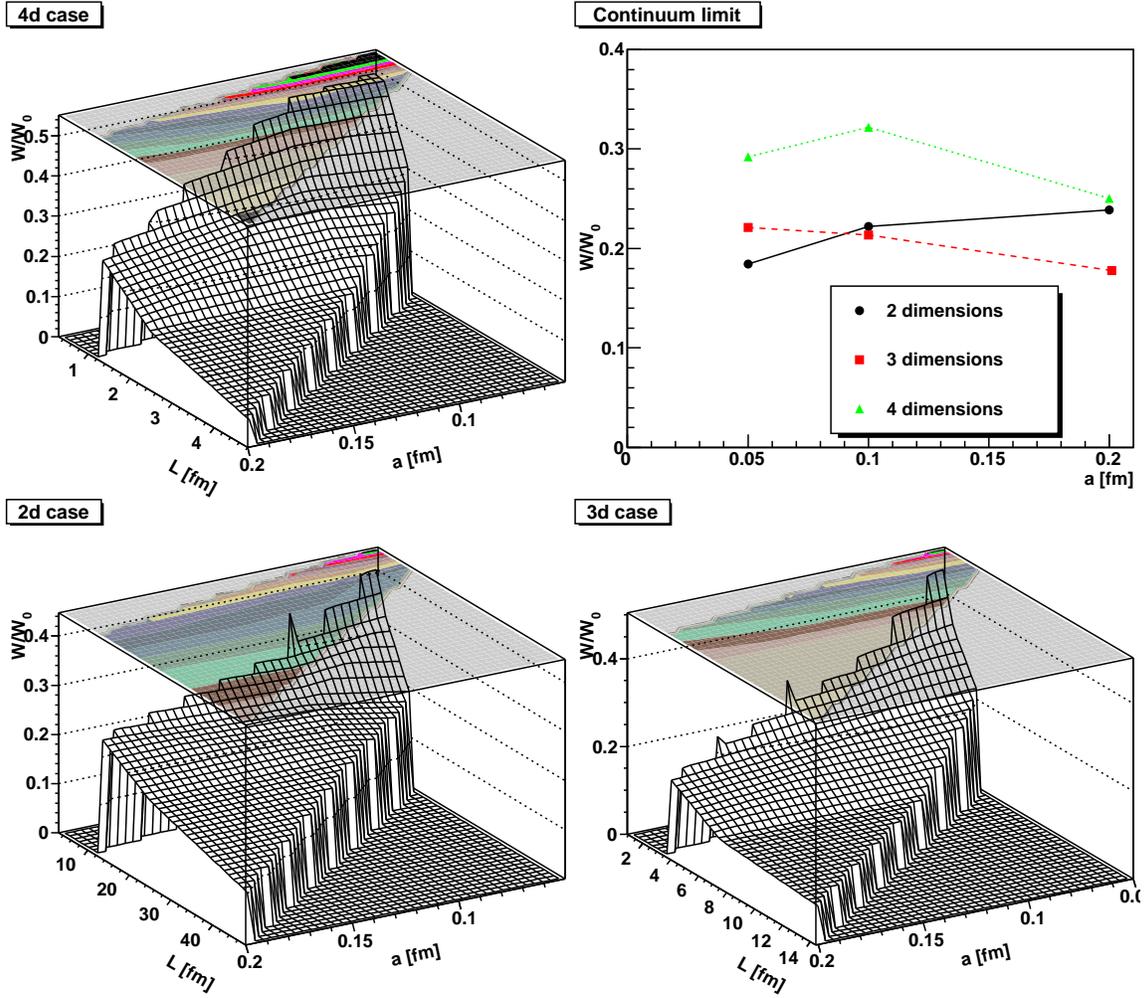}\\
\caption{\label{limits}The value of the normalized free energy density $W(h,k)/W_0$ at $k=0$ in the domain at large $h$, where it becomes constant, as a function of lattice volume and discretization for two (bottom left panel), three (bottom right panel) and four (top left panel) dimensions. The top right panel shows a cut at fixed spatial volume. Statistical error bars are smaller than the symbols. The volume is (12 fm)$^2$ in two dimensions, (3.6 fm)$^3$ in three dimensions, and (1.2 fm)$^4$ in four dimensions.}
\end{figure*}

To investigate whether this is a lattice artifact and/or depends on the dimensionality, the remaining constant of proportionality for a range of lattice sizes and discretizations has been determined \cite{unpublished}, and preliminary results are shown in figure \ref{limits}. At first sight, no qualitative difference is found between different dimensions. It is visible that at fixed lattice spacing the expected bound is less fulfilled the larger the volume. However, when moving towards the continuum at fixed volume the ratio stays, more or less, constant. No final conclusion can be drawn from this, except that the order of limits could be quite important.

Concluding, the numerical results suggest that the qualitative behavior appears to be indeed the one expected from the analytical analysis of section \ref{spw}. Given that this behavior sets in already several order of magnitudes in $h$ below the limit (\ref{hmax}), there is a fair chance that this is a genuine effect and not an artifact of reweighting. However quantitatively the agreement is less convincing, and the bound on the free energy is significantly undersaturated. Since a saturation is only expected asymptotically, this may, or may not, be an artifact of reweighting.  It may, of course, be that configurations saturating the bound are not sufficiently sampled. This may either be due to the importance sampling, or because minimal Landau gauge usually misses the Gribov copies closest to the horizon \cite{Maas:2011se}. Some further considerations will be presented elsewhere \cite{unpublished}.  Furthermore, the dependence on the lattice parameters seen in figure \ref{limits} suggests that the order of limits may be important, which could indicate that non-analyticities may play a role. All of this recommends further investigations, which are currently under way.

\end{document}